\documentclass[11pt]{article} 
\usepackage{ifpdf}
\usepackage{latexsym}
\usepackage{graphics}
\usepackage{epsfig}
\usepackage[dvips]{color}
\usepackage[mathscr]{eucal}
\usepackage{amscd}
\usepackage{amssymb}
\usepackage{amsmath}
\usepackage{amssymb}
\usepackage{amsthm}
\newtheorem{lemma}{Lemma}

\newcommand{\Alex}{\mathbf I}

\newcommand{\Alexflat}{\Alex_0[p,q]}
\newcommand{\Alexflatone}{\Alex_0[x_1,q]}
\newcommand{\volalex}{\mathbf V}
\newcommand{\volflat}{\mathbf {V_0}}
\newcommand{\causet}{\mathbf C} 
\newcommand{\Conef}{\langle C_1\rangle_\eta}
\newcommand{\Ctwof}{\langle C_2\rangle_\eta}
\newcommand{\Cthreef}{\langle C_3\rangle_\eta}
\newcommand{\Ckf}{\langle C_k\rangle_\eta}
\newcommand{\Ckminonexonef}{\langle C_{k-1}(x_1)\rangle_\eta}
\newcommand{\Cone}{\langle C_1\rangle}
\newcommand{\Ctwo}{\langle C_2\rangle}
\newcommand{\Ctwoxone}{\langle C_2(x_1)\rangle}
\newcommand{\Cthree}{\langle C_3 \rangle} 
\newcommand{\Cfour}{\langle C_4 \rangle} 
\newcommand{\Ck}{\langle C_k\rangle}

\newcommand{\Ckplusone}{\langle C_{k+1}\rangle}
\newcommand{\Ckplusonef}{\langle C_{k+1}\rangle_\eta}
\newcommand{\Ckxone}{\langle C_{k}(x_1)\rangle}
\newcommand{\Ckvar}{\Delta C_k}
\newcommand{\Cksq}{\langle C_{k}^2 \rangle}
\newcommand{\Ctwosq}{\langle C_{2}^2 \rangle}

\newcommand{\al}{\mathsf a} 
\newcommand{\be}{\mathsf b} 
\newcommand{\intI}{\mathrm I} 
\newcommand{\intL}{\mathrm L} 
\newcommand{\nno}{\nonumber}
\newcommand{\ff}{{\mathsf f}_0}
\newcommand{\ffc}{{\mathsf f}}
\newcommand{\of}{\mathbf {f}} 
\newcommand{\wzeta}{\zeta_0}
\newcommand{\negs}{\!\!\!\!}
\newcommand{\etalc}{\eta(u,v)}
\title{The Discrete Geometry of a Small Causal Diamond} 
\author{Mriganko Roy, Debdeep Sinha and Sumati Surya \\ 
 Raman Research Institute, Bangalore, India } 

\begin{document}
\maketitle

\begin{abstract} 
  We study the discrete causal set geometry of a small causal diamond in a curved spacetime using
  the average abundance $\Ck$ of $k$-element chains or total orders in the underlying causal set
  $\causet$.  We begin by obtaining the first order curvature corrections to the flat spacetime
  expression for $\Ck$ using Riemann normal coordinates.  For fixed spacetime dimension this allows
  us to find a new expression for the discrete scalar curvature of $\causet$ as well as the 
  time-time component of its Ricci tensor in terms of the $\Ck$.  We also find a new dimension
  estimator for $\causet$ which replaces the flat spacetime Myrheim-Meyer estimator in generic
  curved spacetimes.
\end{abstract}
\section{Introduction} 

That there is a profound relationship between order and Lorentzian geometry has been evident ever
since the work of Malament, Hawking and others \cite{mal,hkm} where they showed the existence of a
bijection between the causal structure, itself a partially ordered set, and the conformal class of
the spacetime metric.  This is one of the main motivations for the causal set approach to quantum
gravity, which assumes that the primitive structure underlying spacetime is a locally finite
partially ordered set, or causal set \cite{blms}. Instead of considering the spacetime metric as the
fundamental dynamical variable in causal set theory (CST) it is the causal structure that one wishes
to ``quantise''.  However, to recover the full spacetime geometry from the causal structure, there
must be a way to  obtain  the spacetime volume or equivalently, the conformal factor.  This is achieved in CST via
the condition of ``local finiteness'' which implies a fundamental spacetime discreteness: underlying
every finite volume  region of spacetime is a  finite cardinality causal set.  Thus, the continuum-discrete correspondence in CST is not exact but approximate, with the
continuum being the approximation of the underlying causal set. 

In order to maintain the relationship between volume and cardinality in all coordinate systems, a
causal set $\causet$ approximated by a spacetime $(M,g)$ is obtained via the following random,
Poisson discretisation of $(M,g)$ \cite{blms,bomhensor}.  Given a fundamental scale $\rho^{-1}$
(which could be the Planck volume) the probability that a spacetime region of volume $V$ contains
$N$ elements of $C$ is given by the Poisson distribution
\begin{equation} 
P_V(N)=\exp^{-\rho V}\frac{(\rho V)^N}{N!}, 
\end{equation}    
for which 
\begin{equation} 
\langle N \rangle = \rho V, 
\end{equation} 
thus establishing the required number to volume correspondence. To give credence to the existence of
a fundamental spacetime discreteness CST moreover requires the following conjecture.  Namely, if a
causal set $\causet$ approximates to a spacetime $(M,g)$, then $(M,g)$ is unique, up to
modifications to it on scales $< \rho^{-1}$.  In other words, this conjectures that {\it all} the
meaningful information about the geometry and topology of $(M,g)$ at scales $>\!\!> \rho^{-1}$ is
contained in the causal set; continuum information below these scales is irrelevant since the
discrete substructure, i.e., the causal set, is fundamental.

From a purely mathematical point of view, this conjecture\footnote{It is also often referred to as a
  ``fundamental theorem'' of CST.}  is very intriguing.  While it has been
verified in several different cases a general proof is still not known, though considerable progress
has been made in this direction \cite{bommeyer,bomnoldus}.  A key question is how to extract
continuum topological and geometric properties from $\causet$ using purely order theoretic
information.  Uniqueness of the approximating spacetime with respect to a {\it given} geometric or
topological property then follows, i.e., any two spacetimes which are approximations to $\causet$
must share this property on scales $> \! \!> \rho^{-1}$.  For example, for a causal set $\causet$
that is approximated by flat spacetime, the Myrheim-Meyer dimension gives a good estimate of the
spacetime dimension \cite{myr,meyer}, while the length of the longest chain or total order between
elements in $\causet$ gives a good estimate of the time-like distance \cite{bg}. An estimator for
spatial distance in this case has also been obtained \cite{rw}.  Additionally, the homology of
spatial hypersurfaces can be constructed from the causal set underlying a globally hyperbolic
spacetime \cite{homone,homtwo}.  A very important recent result is the construction of the scalar
curvature from which  the causal set action is obtained \cite{bd}.

A natural question to ask of the flat spacetime results of \cite{myr,meyer,bg} is how they are
modified in the presence of curvature. While it is true in the continuum that a sufficiently small
neighbourhood of a point is approximately flat, the corrections from curvature can in fact be well
quantified using Riemann normal coordinates in a convex normal neighbourhood of any point.  However,
we run into the following issue in the discrete case: there is at present no known purely order
theoretic definition of a ``small'' neighbourhood of an element in a causal set which corresponds to
a convex normal neighbourhood. At best it may be possible to find approximately flat subsets in a
causal set but it is still unclear how to do this in a systematic way  \cite{lgss}.

This issue however will not be the focus of our current investigation. Instead we consider
only those $N$-element   causal sets $\causet$ which are approximated by  a small causal diamond or Alexandrov
interval $\Alex[p,q]$ in a generic curved spacetime $(M,g)$, where the smallness parameter is given by the
proper time $T$ between the events $p$ and $q$. We expand the metric in 
Riemann normal coordinates (RNC)  about an origin $r=(0,\ldots, 0)$
\begin{equation} 
g_{ab}(x)=\eta_{ab}(0) - \frac{1}{3} x^c x^d R_{acbd}(0) + O(x^3), 
\end{equation} 
where the $O(x)$ correction to the metric vanishes since $\Gamma_{ab}^c(0)=0$.  The RNC has been
used to calculate the volume of a small interval $\Alex[p,q]$ \cite{myr,gs,ks} for which the first
order correction to the volume of $\Alex[p,q]$ due to the effect of curvature occurs at
$O(T^{2+n})$.

The starting point of our analysis is Meyer's work \cite{meyer} in which a general expression was
found for the average abundance $\langle C_k\rangle $ of $k$-chains or $k$-element total orders in a
causal set $\causet_0$ which is approximated by an Alexandroff interval $\Alexflat$ in flat
spacetime.  Following Myrheim, Meyer used this to find a dimension estimator for the dimension of
$\Alexflat$ employing only $\langle C_1\rangle $, the average abundance of elements in $\Alexflat$
and $\langle C_2\rangle $, the average abundance of relations.  Using the RNC expansion and with the
help of \cite{ks} we extend this analysis to the curved spacetime case. We find that
the $\langle C_k\rangle $ in $\Alex[p,q]$ satisfy a recursion relation and depend on the scalar
curvature $R(0)$ and the time-time component of the Ricci tensor $R_{00}(0)$. It is then an easy
exercise to invert these relations and find expressions for $R(0)$ and $R_{00}(0)$ in terms of
$\langle C_{1}\rangle,\Ctwo, \Cthree $ for fixed spacetime dimension. We find that the Myrheim-Meyer
dimension estimator is insufficient to determine dimension in a generic curved spacetime and
construct a new dimension estimator for generic non-flat spacetimes using $\Cone, \Ctwo,\Cthree,
\Cfour$.

As mentioned earlier, the scalar curvature of an element in a causal set was first calculated by
Benincasa and Dowker \cite{bd} using a curved spacetime expression for a non-local D'Alembertian on
a causal set.  They showed that $R(q)$ for an element $q$ in the $\causet$ can be expressed in
terms of the abundance of k-element ``inclusive-intervals'' which are order theoretically very
distinct from $k$-chains. Like $k$-chains they too have a bottom element $e_1 $ and a top element $e_k$, but
unlike $k$-chains, every element in the interval $I[e_1,e_k]$ which is the intersection of the
future of $e_1$ and the past of $e_k$ belongs to the inclusive interval, and has precisely $k-2$
elements satisfying $e_1 \prec e_i \prec e_{k}$.  In contrast, the expression for $R(0)$ that we
find depends only on the abundance of $k$-chains. This may suggest that for manifold-like causal
sets there are hidden relations between these seemingly different order theoretic entities.

The plan of our paper is as follows. In Section \ref{stwo} after first presenting some basic
definitions, we reproduce Meyer's results for the $\Ck$ thus setting the notation that we will use
in the rest of the paper. In Section \ref{sthree} we present the main calculation in the paper,
where we use the RNC to obtain the lowest order curvature correction to $\Ck$. We show that there is a recursion
relation between the coefficients in the expression for the different $\Ck$, but that the form of
the dependence on the $R(0)$ and $R_{00}(0)$ is the same for all $k$. In Section \ref{sfour} for a
fixed spacetime dimension we find expressions for $R(0)$ and $R_{00}(0)$ which depend only on
$\Cone, \Ctwo, \Cthree$. The coefficients in these expressions again have a simple dependence on
dimension. In Section \ref{sfive} we point out that the Myrheim-Meyer dimension estimator is insufficient
in a generic curved spacetime and find a new dimension estimator using $\Cone, \Ctwo, \Cthree, \Cfour$ . An
important question in these calculations is how the error decreases with the sprinkling density
$\rho$ (i.e., the inverse of the volume cut-off) -- the larger $\rho$ is the closer one comes to the
continuum.  In Section \ref{ssix} using the technique developed in \cite{meyer} we show that the
error in $\Ck$ grows as $\rho^{\frac{2k-1}{2}}$ which means that the error in $R(0)$, $R_{00}(0)$
and $n$ goes like $\rho^{-1/2}$, thus going to zero in the continuum limit. We discuss the
implications of our results in Section \ref{sseven} and the questions that need to be addressed in
the future.  Finally, the Appendix contains explicit calculations of the results that appear in the
main body of the paper.

\section{The Abundance of $k$-Chains in Flat Spacetime}  
\label{stwo} 

A $k$-chain in a causal set $\causet$ is a $k$-element total order, i.e., a set of elements $\{e_1,
e_2, \ldots, e_k \}$, $e_i \in \causet$ such that $e_i \prec e_{i+1}$ for all $i$. For any finite
element $\causet$, the number $C_k$ of $k$-chains is therefore invariant of the choice of labeling
of $\causet$. This makes $C_k$ a good observable.  Note that the $e_i$ and $e_{i+1}$ need not be
``linked'', i.e., there could exist an element $e \in \causet $ such that $e_i \prec e \prec
e_{i+1}$. Moreover, $e_1 \prec e \prec e_{k}$ does not imply that $e$ belongs to the $k$-chain.  In
contrast, a $k$-inclusive interval is defined as $I_k[e_1,e_{k}]= {\mathrm {Future}}(e_1) \cap
{\mathrm {Past}}(e_{k}) $ \cite{bd}. Along with the elements $e_1,e_k$ it also contains precisely
$k$-elements.  However, every $e \in \causet$ such that $e_1 \prec e \prec e_{k}$ belongs to
$I_k[e_1,e_k]$, which means that the order theoretic structure of a $k$-chain is very different from
that of a $k$-inclusive interval.  The fact that one can express the discrete scalar curvature 
both in terms of the abundance of the inclusive intervals as shown by Benincasa and Dowker \cite{bd}
and in terms of the abundance of $k$-chains  as we will show in Section \ref{sfour} thus suggests a hidden
connection between the two.

We now reproduce Meyer's results in $n$-dimensional flat spacetime $(M_0,\eta)$ using
notation that we will find convenient in the curved spacetime generalisation. Let $p, q \in M_0$
such that $p=(-T/2,0,\ldots, 0)$ and $q=(T/2, 0, \ldots, 0)$. For a causal set $\causet_0$ that is
approximated by $\Alexflat$ for a given sprinkling density $\rho$,  the average abundance of elements, or 
$1$-chains  $\Conef$ is given by
\begin{eqnarray} 
\label{volumeflat} 
\Conef &=& \rho \volflat =\rho\negs \int\limits_{\Alexflat} \negs dx_1\nno \\ 
&=&  2 \rho \int\limits_{0}^{\frac{T}{2}} dt 
\int\limits_0^{\frac{T}{2}-t}  dr \,\, r^{n-2}  \int d\Omega_{n-2} = \rho\frac{2
  A_{n-2}}{n(n-1)}\biggl(\frac{T}{2}\biggr)^n = \rho \,\wzeta T^n,   
\end{eqnarray} 
where $A_{n-2}$ is the volume of the unit $(n-2)$ sphere $S^{n-2}$.    
Next, the average number of  $2$-chains or relations in $\Alexflat$ is given by  the probability
of there being a pair of elements $x_1,x_2 \in \Alexflat$ such that $x_1 \prec x_2$, or 
\begin{equation} 
\Ctwof=\rho^2 \negs \int\limits_{\Alexflat}\negs dx_1 \negs \int\limits_{J^+(x_1) \cap \Alexflat}\negs
\negs \negs dx_2  
\end{equation}     
Recognising that  the integral over $dx_2$  is simply the volume of the smaller interval
$\Alexflatone$ and using Eq(\ref{volumeflat})  
\begin{equation} 
\label{ctwoflat} 
\Ctwof= \rho^2 \frac{2 A_{n-2}}{2^nn(n-1)} \int\limits_{\Alexflat} dx_1 T_1^n = \rho^2 \volflat^2
\frac{\Gamma(n+1) \Gamma(\frac{n}{2})}{4 \Gamma(\frac{3n}{2})}  
\end{equation}
Meyer was able to similarly use the nested integral expression for $\Ckf$  
\begin{equation} 
\label{recur} 
\Ckf=\rho^k\negs \int\limits_{\Alexflat}\negs  dx_1  \negs \int\limits_{J^+(x_1) \cap \Alexflat}\negs
dx_2 \ldots \negs \int\limits_{J^+(x_{k-1}) \cap \Alexflat} \negs dx_k =\rho^k \negs 
\int\limits_{\Alexflat} \negs  dx_1  \Ckminonexonef
\end{equation} 
to find by induction the general form  
\begin{equation} 
\Ckf=\rho^k\chi_k \volflat^k=\rho^k\zeta_k T^{kn},
\end{equation}  
where 
\begin{equation} 
\chi_k\equiv \frac{1}{k} \biggl(\frac{\Gamma(n+1)}{2}\biggr)^{k-1}
\frac{\Gamma(\frac{n}{2})\Gamma(n)}{\Gamma(\frac{kn}{2})\Gamma(\frac{(k+1)n}{2})}, \quad \zeta_k
\equiv \biggl(\frac{2
  A_{n-2}}{2^nn(n-1)}\biggr)^k \chi_k = {\wzeta}^k \chi_k ,  
\end{equation} 
with $\wzeta$ defined as in Eqn (\ref{volumeflat}). Note in particular that $\chi_1=1$.  We will
find it useful to express $\Ckf$ as
\begin{equation}
\label{rrecur} 
\Ckf=\rho^k\, \zeta_{k-1} \negs \int\limits_{\Alexflat} \negs dx_1 T_1^{(k-1)n} = \rho^k \zeta_{k-1} \intI_1((k-1)n) 
\end{equation}  
where $\intI_1(m)$ is evaluated in Eqn (\ref{intone}) of the Appendix.  As discussed above, the
average number of chains in a finite element causal set $\causet$ is itself a covariant
observable. In particular, the distribution of the abundance of $k$-chains as a function of $k$ in a
finite element causal set can be compared with the distribution of $\Ckf$; if the two distributions
agree, it is an indication that the $\causet$ may be well approximated by flat spacetime and is
therefore manifold-like. A similar comparison using $k$-inclusive intervals was found to be useful
in determining flat spacetime behaviour in a model of 2d causal set quantum gravity \cite{ss}.  It
is therefore important to find a generalisation of $\Ckf$ to curved spacetime.

Meyer obtained a dimension estimator from $\Ckf$ by observing that the ratio 
\begin{equation} 
\label{dimension} 
\ff(n) \equiv  \frac{\Ctwof}{\Conef^2} = \frac{\Gamma(n+1) \Gamma(\frac{n}{2})}{4 \Gamma(\frac{3n}{2})} 
\end{equation} 
is only a function of $n$. Thus, one has an expression for the dimension which depends only on
order-theoretic information in the causal set.   Indeed, $\ff(n)$ is one-half  of  Myrheim's
ordering fraction 
\begin{equation} 
\of(\causet) \equiv  R \binom{N}{2}^{-1} \approx \frac{2R}{N^2} 
\end{equation}  
where $R=\langle C_2\rangle $ is the number of relations and $N = \langle C_1\rangle $. In 2
spacetime dimensions for example, $\of(2)=1/2$, i.e., the inverse of the spacetime
dimension.  In Section \ref{sfive} we will show that Eqn (\ref{dimension}) does not suffice in curved spacetime and
there is need for a new dimension estimator. 


\section{The Abundance of $k$-Chains in a Small Causal Diamond in Curved Spacetime}  
\label{sthree}
\begin{figure}[ht] 
\centering \resizebox{2.0in}{!}{\includegraphics{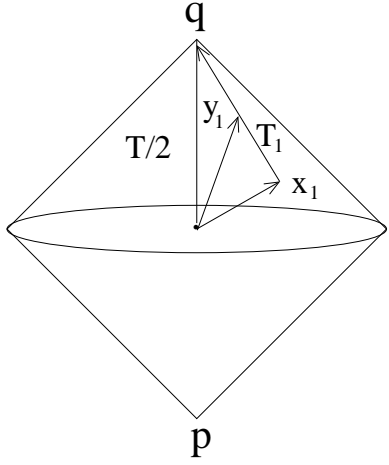}}
\vspace{0.5cm}
\caption{{\small An Alexandroff interval  $\Alex[p,q]$ in flat spacetime. $T_1$ is the proper time
    between the events $x_1$ and $q$ and $y_1$ is the midpoint of $\Alex[x_1,q]$. }} \label{ctwo.fig}
\end{figure}

The RNC expansion to order $T^2$ gives an expression for $\Cone$ in $\Alex[p,q]$ \cite{myr,gs}
\begin{equation} 
\label{volume} 
\Cone = \rho\volalex = \rho\negs \int\limits_{\Alex[p,q]} \sqrt{-g_1} \,\,dx_1 =\rho  \volflat\biggl( 1+ \alpha_1 R(0) T^2 + \beta_1 R_{00}(0) T^2 \biggr)
\end{equation}    
where 
\begin{equation}
\alpha_1 = -\frac{n}{24(n+1)(n+2)}, \qquad \beta_1= \frac{n}{24(n+1)}, 
\end{equation} 
and which  uses the RNC expansion 
\begin{equation} 
\sqrt{-g_1} = 1 - \frac{1}{6} x_1^\mu x_1^\nu R_{\mu\nu}(0) + O(x^3). 
\end{equation} 
Now, the average number of 2-chains or relations is given by the similar generalisation 
\begin{equation}
\Ctwo= \rho^2\negs \int\limits_{\Alex[p,q]} \sqrt{-g_1} \,\, dx_1 \negs \negs \int\limits_{J^+(x_1) \cap \Alex[p,q]}
\negs \negs\sqrt{-g_2} \, \, dx_2 = \rho^2 \negs \int\limits_{\Alex[p,q]} \negs \sqrt{-g_1} \, \,
dx_1 \volalex_1, 
\end{equation} 
where $\volalex_1$ denotes the volume of the region $J^+(x_1) \cap \Alex[p,q]$.    
Using the covariant form of Eqn (\ref{volume}) we see that 
\begin{eqnarray}
\label{ctwo} 
\Ctwo & = &\rho^2 \zeta_1 \int\limits_{\Alex[p,q]} dx_1 T_1^n \biggl( 1-\frac{1}{6} x_1^\mu x_1^\nu R_{\mu\nu}(0)+ \alpha_1
T_1^2 R(y_1) + \beta_1 T_1^\mu T_1^\nu R_{\mu\nu}(y_1)\biggr)+ O(T^{n+3}) \nonumber \\ 
&=& \rho^2 \zeta_1 \biggl[ \int\limits_{\Alex[p,q]} dx_1 T_1^n  \nonumber \\ 
&&  + \int\limits_{\Alexflat} dx_1 T_1^n \biggl( -\frac{1}{6} x_1^\mu x_1^\nu R_{\mu\nu}(0)+ \alpha_1 T_1^2 R(y_1) + \beta_1 T_1^\mu T_1^\nu
R_{\mu\nu}(y_1)\biggr)\biggr] + O(T^{n+3}), \nonumber \\
\end{eqnarray} 
where we have split the integral in the manner of \cite{myr} : the first is a {\it flat spacetime}
integral over the curved spacetime interval $\Alex[p,q]$, whereas the second is the contribution
from the curvature terms over the flat spacetime interval $\Alexflat$. $y_1 $ is the midpoint of the interval
$\Alex[x_1,q]$  as shown in   Fig \ref{ctwo.fig}. 
Using the light-cone coordinates $u= t - r$ , $ v= t+r$  we see from Fig \ref{ctwo.fig} that  
\begin{equation} 
T_1^\mu = \biggl(\frac{T}{2}\biggr)^\mu -x_1^\mu, \qquad T_1
=\sqrt{\biggl(\frac{T}{2}-u_1\biggr)\biggl(\frac{T}{2}-v_1\biggr) }. 
\end{equation} 
Thus, 
\begin{equation} 
T_1^\mu T_1^\nu R_{\mu\nu}(y_1) =  \frac{T^2}{4}R_{00}(y_1) - x_1^\mu T R_{0\mu}(y_1) +  x_1^\mu
x_1^\nu  R_{\mu\nu}(y_1).  
\end{equation} 
The first integral was
evaluated in \cite{ks} and shown to be of the form
\begin{equation}
\label{MMI}  
\int\limits_{\Alex[p,q]} dx_1 T_1^m= \int\limits_{\Alexflat} dx_1T_1^m\biggl(1+\frac{T^2}{24} R_{00}(0)
\biggr)  
\end{equation} 
for any non-negative integer $m$.  Moreover, as can be readily seen, to order $T^2$,
$R_{\mu\nu}(y_1)$ can be replaced with $R_{\mu\nu}(0)$.  We can moreover simplify the expressions in
$\Ctwo$ substantially by using the symmetries of $\Alexflat$.   Expanding the term  
\begin{equation} 
\int\limits_{\Alexflat}  dx _1 x_1^\mu x_1^\nu R_{\mu\nu}(0) T_1^m =   \int\limits_{\Alexflat}  dx _1 t_1^2
R_{00}(0) T_1^m  +2  \int\limits_{\Alexflat}  dx _1 t x_1^i R_{0i}(0) T_1^m + \int\limits_{\Alexflat}  dx _1 x_1^i
x_1^j R_{ij}(0) T_1^m, 
\end{equation}  
for $m$ a positive integer,  we see that the cross-terms do not contribute, so that we are left with 
\begin{equation} 
\int\limits_{\Alexflat}  dx _1 x_1^\mu x_1^\nu R_{\mu\nu}(0) T_1^m =   R_{00}(0)  \int\limits_{\Alexflat}  dx _1 t_1^2
T_1^m + \sum\limits_{i=1}^{n-1} R_{ii}(0) \int\limits_{\Alexflat}  dx _1 (x_1^i)^2 T_1^m,   
\end{equation} 
since the last integral is independent of the spatial direction $i$, due to the  symmetry of
$\Alexflat$.  

Gathering  the coefficients of $R(0)$ and
$R_{00}(0)$
\begin{eqnarray} 
\Ctwo &=& \rho^2 \zeta_1 \biggl[ \intI_1(n) + R(0) \biggl[(\beta_1 - \frac{1}{6}) \intI_2(n) + \alpha_1
\intI_1(n+2) \biggr] + \nno \\ 
&& R_{00}(0) \biggl[ \frac{T^2}{24}\intI_1(n) + (\beta_1 - \frac{1}{6}) 
\intI_2(n) -\frac{1}{6} \intI_3(n) + \beta_1 \intI_4(n) \biggr] \biggr], 
\end{eqnarray} 
where we have used $\sum\limits_{i=1}^{n-1} R_{ii}(0)=R_{00}(0) + R(0)$ and  we define the general class of integrals 
\begin{eqnarray} 
\intI_1(m)=\int\limits_{\Alexflat} dx_1 T_1^m,   &\quad  & \intI_2(m) = \int\limits_{\Alexflat} dx_1T_1^m r_1^2 \cos^2\theta_1 \nonumber \\ 
\intI_3(m)= \int\limits_{\Alexflat}  dx_1 t_1^2 T_1^m, &\quad & \intI_4(m) =\int\limits_{\Alexflat}  dx_1 \biggl(\frac{T}{2}-t_1 \biggr)^2 T_1^m.  
\end{eqnarray} 
for non-negative integers $m$. 
These integrals have been evaluated in the Appendix. Using the terminology defined therein, we will
find it useful to re-express the last three integrals in terms of the first 
\begin{equation} 
\intI_2(m)=f_2(m)\intI_1(m)  T^2, \quad \intI_3(m)=f_3(m)\intI_1(m)  T^2, \quad \intI_4(m)=f_4(m)
\intI_1(m) T^2 \nno 
\end{equation} 
and 
\begin{equation}
\intI_1(m+2)=g_1(m) \intI_1(m) T^2, \nno  
\end{equation} 
where  $g_1(m), f_2(m), f_3(m)$ and $f_4(m)$ are defined in Eqn (\ref{gfs}) of the Appendix . 
Given Eqn. (\ref{rrecur})  we find that 
\begin{equation} 
\Ctwo=\Ctwof\biggl[1+ T^2 \alpha_2 R(0)+ T^2 \beta_2 R_{00}(0) \biggr] 
\end{equation} 
where 
\begin{eqnarray} 
\alpha_2 &=& (\beta_1 - \frac{1}{6}) f_2(n) + \alpha_1
g_1(n) \nno  \\
\beta_2&=& \frac{1}{24} + (\beta_1 - \frac{1}{6}) f_2(n) -\frac{1}{6} f_3(n) + \beta_1 f_4(n).  
\end{eqnarray} 
We find that  
\begin{eqnarray} 
\alpha_2 &=&-\frac{4n}{24(2n+2)(3n+2)}\nonumber \\ 
\beta_2 &=& \frac{4n}{24(3n+2)}.
\end{eqnarray}

We can go one step further and calculate 
\begin{eqnarray} 
\label{cthree} 
\Cthree &=& \rho  \int dx_1 \, \, \sqrt{-g_1}  \Ctwoxone \nno \\ 
&=& \rho^3 \zeta_2   \biggl[ \intI_1(2n) + R(0) \biggl((\beta_2 - \frac{1}{6}) \intI_2(2n) + \alpha_2
\intI_1(2n+2) \biggr) + \nno \\ 
&& R_{00}(0) \biggl(\frac{T^2}{24}\intI_1(2n) + (\beta_2 - \frac{1}{6}) 
\intI_2(2n) -\frac{1}{6} \intI_3(2n) + \beta_2 \intI_4(2n) \biggr)\biggr], \nno \\ 
&=& \Cthreef \biggl[1+ T^2 \alpha_3 R(0)+ T^2 \beta_3 R_{00}(0) \biggr]
\end{eqnarray} 
where again we have used Eqn (\ref{rrecur}) and 
\begin{eqnarray} 
\alpha_3 &=& (\beta_2 - \frac{1}{6}) f_2(2n) + \alpha_2
g_1(2n)= -\frac{6n}{24 (3n+2)(4n+2)}\nonumber \\ 
\beta_3 &=& \frac{1}{24} + (\beta_2 - \frac{1}{6}) f_2(2n) -\frac{1}{6} f_3(2n) + \beta_2 f_4(2n) =
\frac{6n}{24(4n+2)}. 
\end{eqnarray} 
 This suggests an
iterative formula
\begin{eqnarray} 
\alpha_{k+1} & = &  (\beta_k - \frac{1}{6}) f_2(kn) + \alpha_k
g_1(kn) \nno \\ 
\beta_{k+1} &= & \frac{1}{24} + (\beta_k - \frac{1}{6}) f_2(kn) -\frac{1}{6} f_3(kn) + \beta_k
f_4(kn), 
\end{eqnarray} 
with 
\begin{eqnarray} 
\label{alphabetak} 
\alpha_k &=& -\frac{nk}{12(kn+2)((k+1)n+2)} \nno \\
\beta_k &=& \frac{nk}{12((k+1)n+2)}.  
\end{eqnarray} 

\begin{lemma}  To the lowest order correction in the flat spacetime expression, the average number of $k$-element
chains in a small causal diamond is  
\begin{equation} 
\label{ck} 
\Ck= \Ckf \biggl[1+ T^2 \alpha_{k} R(0)+ T^2 \beta_{k} R_{00}(0) \biggr] + O(T^{kn+3}), 
\end{equation} 
where $\alpha_k$ and $\beta_k $ are given by Eqn (\ref{alphabetak}). 
\end{lemma} 

\noindent {\bf Proof: } 
We will prove this inductively. We have already shown it for  $k=2$.  Now, let us assume 
$\Ck$ is of the form Eqn. (\ref{ck}).  Just as  in the flat spacetime case,  one has nested
integrals so that 
\begin{eqnarray}
\label{ckplusone}  
\Ckplusone\!\!\! &= &\!\! \!\!\rho \int\limits_{\Alex[p,q]}\!\!\!  dx_1 \sqrt{-g_1} \Ckxone \nno \\ 
&=& \!\!\!\!\rho^{k+1}\zeta_k\!\!\!\! \int\limits_{\Alexflat}\!\!\! \!\!dx_1  T_1^{kn} \biggl( 1+
\frac{T^2}{24}R_{00}(0) -\frac{1}{6} x_1^\mu
x_1^\nu R_{\mu\nu}(0)+ \alpha_k 
T_1^2 R(0) + \beta_kT_1^\mu T_1^\nu R_{\mu\nu}(0)\biggr)\nno \\ 
&& \,\,\,\,+ \, \, \,O(T^{kn+3})  
\end{eqnarray} 
where we have used Eqn (\ref{MMI}) to reduce the integral over $\Alex[p,q]$ to one over $\Alexflat$
to order $O(T^{kn+2})$.   Using the integrals $\intI_{1,2,3,4}(m)$ from the Appendix, we can reduce
this to 
\begin{eqnarray} 
\Ckplusone &=& \rho^{k+1} \zeta_k \biggl[\intI_1(kn)  + R(0)\biggl(  (\beta_{k} - \frac{1}{6}) \intI_2(kn) +
\alpha_k \intI_1(kn+2)   \biggr) + \nno \\ 
&& R_{00}(0) \biggl( \frac{T^2}{24}\intI_1(kn) + (\beta_k - \frac{1}{6}) 
\intI_2(kn) -\frac{1}{6} \intI_3(kn) + \beta_k \intI_4(kn) \biggr) \biggr], \nno \\ 
&=& \Ckplusonef \biggl[1  +T^2  R(0)\biggl(  (\beta_{k} - \frac{1}{6}) f_2(kn) +
\alpha_k g_1(kn+2)   \biggr) + \nno \\ 
&& T^2 R_{00}(0) \biggl( \frac{1}{24}+ (\beta_k - \frac{1}{6}) 
f_2(kn) -\frac{1}{6} f_3(kn) + \beta_k f_4(kn) \biggr) \biggr].  \nno \\ 
\end{eqnarray} 
Writing it in the form Eq (\ref{ck}), with 
\begin{eqnarray} 
\alpha_{k+1} &=&   (\beta_{k} - \frac{1}{6}) f_2(kn) + \alpha_{k}
g_1(kn+2)\nno \\ 
\beta_{k+1}&=&   \frac{1}{24} + (\beta_{k} - \frac{1}{6}) f_2(kn) -\frac{1}{6} f_3(kn) + \beta_{k} f_4(kn).  
\end{eqnarray}
we find the desired form for $\Ckplusone$.   It then follows from the expressions for $g_1(m),f_2(m),
f_3(m)$ and $f_4(m)$ (Eqns (\ref{ione}),(\ref{itwo}),(\ref{ithree}),(\ref{ifour}))  that 
\begin{eqnarray} 
\alpha_{k+1} & = & -\frac{n(k+1)}{12((k+1)n+2)((k+2)n+2)}  \nno \\ 
\beta_{k+1} &= & \frac{n(k+1)}{12((k+2)n+2)}.
\end{eqnarray} 

\hfill $\qed$ 

\section{Scalar Curvature from the Abundance of $k$-Chains}  
\label{sfour} 

For a fixed $n$  $\Ck$ contains three unknowns, $T, R(0)$ and $R_{00}(0)$.  Thus, we need at least
three values of $k$ in order to determine $R(0)$. For each $\Ck$ the lowest order correction due to
curvature is $O(T^{kn+2})$. Hence, as in the flat spacetime calculation of the Myrheim-Meyer
dimension Eqn (\ref{dimension}) we must take appropriate powers of $\Cone,\Ctwo$ and $ \Cthree$ to
be able to compare their lowest order corrections. Defining 
\begin{equation} 
Q_k\equiv \biggl(\frac{\Ck}{\rho^k\zeta_k }\biggr)^{3/k} =\frac{1}{\wzeta^3}\biggl(\frac{\Ck}{\rho^k\chi_k }\biggr)^{3/k} 
\end{equation} 
for $k=1,2,3$ 
\begin{eqnarray} 
Q_1&=& T^{3n} \biggl( 1+ 3 \alpha_1 R(0) T^2 + 3 \beta_1 R_{00}(0)T^2
\biggr) + O(T^{3n+2})\label{qone} \\ 
Q_2&=& T^{3n} \biggl( 1+ \frac{3}{2} \alpha_2 R(0)
T^2 + \frac{3}{2}  \beta_2 R_{00}(0)T^2 \biggr) + O(T^{3n+2})\label{qtwo} \\ 
Q_3&=& T^{3n} \biggl( 1+ \alpha_3 R(0) T^2 +  \beta_3 R_{00}(0)T^2 \biggr) +
O(T^{3n+2}). \label{qthree}   
\end{eqnarray}     
Thus, the $Q_k$ are independent of the sprinkling density $\rho$ and hence can be used to construct
continuum geometric parameters.  It is useful to gather a few identities and definitions before we
proceed:
\begin{eqnarray} 
\label{moreidentities}
\beta_k \alpha_k^{-1} & = & -(kn+2)  \nno \\ 
\Psi_k&\equiv& \alpha_k \beta_{k+1} - \alpha_{k+1}\beta_k = -n \alpha_k\alpha_{k+1}\nno \\ 
\Phi_k&\equiv& \frac{k}{k+1}\beta_{k+1}-\beta_{k}\nno \\  
\Theta_k &=&\frac{k}{k+1}\alpha_{k+1}-\alpha_{k} \nno \\ 
 K_k &\equiv& ((k+1)n+2)Q_k, \nno \\ 
  J_k &\equiv& (kn+2) K_k
\end{eqnarray} 

We eliminate the $R_{00}(0)$ term from Eqn (\ref{qone}) and (\ref{qtwo}) and subsequently from  Eqn
(\ref{qtwo}) and (\ref{qthree}) to get the pair of equations 
\begin{eqnarray} 
\biggl(\frac{\beta_2}{2}Q_1-\beta_1Q_2\biggr) T^{-3n}&=& \Phi_1  +
\frac{3}{2} \Psi_1T^2 R(0) \label{four} \\
\biggl(\frac{2\beta_3}{3}Q_2-\beta_2Q_3\biggr) T^{-3n}&=& \Phi_2 +
\Psi_2T^2 R(0) \label{five}.  
\end{eqnarray} 
Since both $R(0)$ and $T$ are unknowns,  we first eliminate the $R(0)T^2$ term: 
\begin{equation}
 \biggl(\frac{\beta_2}{3}\Psi_2Q_1-\frac{2}{3}(\beta_1 \Psi_2 - \beta_3 \Psi_1) Q_2+ \beta_2\Psi_1Q_3\biggr) T^{-3n}
= \frac{2}{3} \Psi_2\Phi_1 - \Psi_1 \Phi_2. 
\end{equation}  
We find after some algebraic manipulation that 
\begin{equation} 
\label{T}
T^{3n}=\frac{1}{2n^2}\biggl(J_1-2J_2+J_3 \biggr).   
\end{equation} 
with the $J_i$'s given by  Eqn. (\ref{moreidentities}).  We thus obtain the expression for the
scalar curvature
\begin{equation} 
\label{scalarcurvature} 
R(0)= -\frac{2(n+2)(2n+2)(3n+2)}{n^3 T^{3n+2}}(K_1-2K_2+K_3) 
\end{equation} 
or more explicitly 
\begin{equation} 
\label{R}
R(0)= 
-{2(n+2)(2n+2)(3n+2)} 2^{\frac{3n+2}{3n}} n^{\frac{4}{3n}-1} \frac{ (K_1-2K_2+K_3)}{\,\,\,\,\,(J_1 - 2J_2 +
  J_3)^{\frac{3n+2}{3n}}}.
\end{equation}

We may  additionally solve for $R_{00}(0)$ by eliminating $R(0)$ from Eqn (\ref{qone}),
(\ref{qtwo})(\ref{qthree}): 
\begin{eqnarray} 
\biggl(\frac{\alpha_2}{2}Q_1-\alpha_1Q_2\biggr) T^{-3n}&=& \Theta_1  -
\frac{3}{2} \Psi_1T^2 R(0) \label{seven} \\
\biggl(\frac{2\alpha_3}{3}Q_2-\alpha_2Q_3\biggr) T^{-3n}&=& \Theta_2 - 
\Psi_2T^2 R(0) \label{eight}.  
\end{eqnarray}     
Either equation along with Eqn (\ref{T}) gives
\begin{equation}
\label{Roo}
R_{00}(0)=  -\frac{4(2n+2)(3n+2)}{n^3 T^{3n+2}}((n+2)Q_1-(5n+4)Q_2+(4n+2)Q_3). 
\end{equation}

As a check, let  us consider the case $R_{00}(0)=0$  so that from Eqn
(\ref{qone})-(\ref{qthree}) 
we see that
\begin{eqnarray} 
R(0) >0  &\Rightarrow&  Q_1 < Q_2 < Q_3  \nno \\ 
R(0)<0 &\Rightarrow&  Q_1 > Q_2 > Q_3 \nno \\ 
R(0)=0  &\Rightarrow&  Q_1 = Q_2 = Q_3=T^{3n}.  
\end{eqnarray} 
Moreover, 
\begin{equation}
K_1-2K_2+K_3  = - \frac{ n^3}{2(n+2)(2n+2)(3n+2)} R(0)T^{3n+2}  
\end{equation} 
which therefore has the opposite sign to  $R(0)$ and is zero when $R(0)=0$. 

As a further check, we note that if both $R_{00}(0)=0$ and $R(0)=0$, $Q_1 = Q_2 = Q_3=T^{3n}$ so that
not only is $K_1-2K_2+K_3=(2n+2)Q_1-2(3n+2)Q_2+ (4n+2)Q_3=0$, but also
$(n+2)Q_1-(5n+4)Q_2+(4n+2)Q_3=0$ which appears in Eqn (\ref{Roo}).   

\section{A New Dimension Estimator for  Curved Spacetime}  
\label{sfive} 

As one can guess by now, the ordering fraction or equivalently the function $\ffc(n)$ in curved spacetime clearly involves
the curvature contribution non-trivially. Expanding to order $T^2$ the curved spacetime version of
the Myrheim-Meyer dimension estimator is 
\begin{eqnarray}
\label{curvedff}  
\ffc(n) &=&   \frac{\Ctwo}{{\Cone}^2} \nno \\ 
& =&  \ff(n) \biggl(1+ T^2 (\alpha_2-2\alpha_1) R(0) + T^2 (\beta_2-2\beta_1) 
  R_{00}(0)\biggr) + O(T^3) \nno \\ 
&=&   \ff(n) \biggl(1+ \frac{n^2T^2}{12(n+1)(3n+2)} \biggl(\frac{2}{(n+2)} R(0) -
R_{00}(0)\biggr)\biggr) +O(T^3). 
\end{eqnarray} 
In the special case that $R(0)=R_{00}(0)=0$,  $\ffc(n) \approx \ff(n)$ up to order $T^2$. For a
generic spacetime $\ffc(n)$ it is however insufficient as a dimension estimator and we
must find a replacement. Given that along with $n$ there are   $4$ unknowns to be solved in terms of
the $\Ck$, the simplest way to do so is to  include $\Cfour$ in our analysis. 
  
We define \begin{equation} 
S_k=\biggl(\Ck/\rho^k \zeta_k\biggr)^{4/k}
\end{equation} 
analogous to the $Q_i$ in the previous section, with $k=1,2,3,4$ which again is independent of the
sprinkling density.  To the lowest order correction we have the  four equations 
 \begin{eqnarray} 
S_1 &=& T^{4n}\biggl( 1+ 4 \alpha_1 R(0) T^2 + 4 \beta_1 R_{00}(0) T^2 \biggr ) \nno \\ 
S_2 &=& T^{4n}\biggl( 1+ 2 \alpha_2 R(0) T^2 + 2 \beta_2 R_{00}(0) T^2 \biggr ) \nno \\ 
S_3 &=& T^{4n}\biggl( 1+ \frac{4}{3} \alpha_3 R(0) T^2 + \frac{4}{3} \beta_3 R_{00}(0) T^2 \biggr ) \nno \\ 
S_4 &=& T^{4n}\biggl( 1+  \alpha_4 R(0) T^2 + \beta_4 R_{00}(0) T^2 \biggr ) 
\end{eqnarray}  
Eliminating $R(0)T^2$ from the above we get 
\begin{eqnarray}
\biggl(\frac{1}{2} \alpha_2 S_1 -\alpha_1S_2 \biggr)&=&T^{4n}  \biggl( \Theta_1 -2 \Psi_1
R_{00}T^2\biggr) \nno \\  
\biggl(\frac{2}{3} \alpha_3 S_2 -\alpha_2S_3 \biggr) &=&T^{4n}  \biggl( \Theta_2 -\frac{4}{3} \Psi_2
R_{00}T^2\biggr) \nno \\  
\biggl(\frac{3}{4} \alpha_4 S_3 -\alpha_3S_4 \biggr)&=& T^{4n}  \biggl( \Theta_3 -\Psi_3
R_{00}T^2\biggr)   
\end{eqnarray} 
from which we may eliminate $R_{00}(0)T^2$ to get  
\begin{eqnarray} 
\frac{4}{3} \biggl(\frac{1}{2} \alpha_2 S_1 -\alpha_1S_2 \biggr)\Psi_2 - 2 \biggl(\frac{2}{3}
\alpha_3 S_2 -\alpha_2S_3 \biggr) \Psi_1 &=& T^{4n} \biggl( \frac{4}{3}\Theta_1 \Psi_2 - 2 \Theta_2
\Psi_1 \biggr) \nno \\ 
\frac{3}{4} \biggl(\frac{2}{3} \alpha_3 S_2 -\alpha_2S_3 \biggr) \Psi_3 - \biggl(\frac{3}{4}
\alpha_4 S_3 -\alpha_3S_4 \biggr) \Psi_2 &=& T^{4n} \biggl( \frac{3}{4}\Theta_2 \Psi_3 - \Theta_3
\Psi_2 \biggr).  \nno 
\end{eqnarray}  
This gives us an expression for $T^{4n}$ (which we check reduces to Eqn(\ref{T})).    
After some algebra this gives the following implicit form for the dimension:   
\begin{equation}
\label{edim} 
(n+2)(2n+2)S_1 -3 (2n+2)(3n+2)S_2 + 3 (3n+2)(4n+2)S_3 -(4n+2)(5n+2)S_4=0. 
\end{equation} 
Importantly, in the absence of curvature the $S_k$ are all equal and the left-hand side is identically
zero. Hence this cannot be a replacement for the Myrheim-Meyer dimension in flat spacetime. 
Using $U_k=(k+2)((k+1)n+2)S_k$ we may write the expression more succinctly as 
\begin{equation}
\label{U}
U_1-3U_2+3U_3-U_4=0.   
\end{equation} 
It is interesting to note the appearance of the binomial coefficients $(-1)^k \binom{r-1}{k}$ for $r=4$
in the above expression for the dimension estimator, as well as in the expressions for $R(0)$ and
$T$ for $r=3$.    

Since the $S_k$ themselves explicitly contain dimension information via the $\zeta_k$  it
is more useful to expand the expression to 
\begin{eqnarray} 
(n+2)(2n+2)\biggl(\frac{\Cone}{\chi_1}\biggr)^4  - 3 (2n+2)(3n+2)
\biggl(\frac{\Ctwo}{\chi_2}\biggr)^2 & & \nno \\ 
  + 3 (3n+2)(4n+2) \biggl(\frac{\Cthree}{\chi_3}\biggr)^{4/3}   -(4n+2)(5n+2)
  \biggl(\frac{\Cfour}{\chi_4}\biggr)  &= & 0 
\end{eqnarray} 
or defining 
\begin{equation} 
\quad \omega_k \equiv (-)^{k-1}\binom{3}{k-1}\frac{(kn+2)((k+1)n+2)}{\chi_k^{4/k}}
\end{equation} 
we get our final expression for the dimension estimator in curved spacetime  
\begin{equation} 
\label{dim}
\sum\limits_{k=1}^4 \omega_k(n) \Ck^{4/k} = 0.
\end{equation} 

\section{Calculating the Errors} 
\label{ssix} 

We expect that as the sprinkling density $\rho$ increases, our curvature and dimension estimators
should do a better job of reproducing the continuum results.  While the geometric parameters
themselves do not depend on $\rho$, it is clear that the error will.  The (rms) error $\delta C_k =
\sqrt{ \Delta C_k}= \sqrt{\Cksq-\Ck^2}$, where $\Ckvar$ is the variance.  We follow the analysis in
\cite{meyer} to find the dependence of $\Ckvar$ on $\rho$ in the RNC. Unlike the flat spacetime
case however, it is not the continuum volume that must be increased to improve accuracy  since
this region must still be ``small''  for the RNC to be valid. Instead, it is the sprinkling
density $\rho$ that should be increased for reducing the error. 

Let us begin with $k=2$, so that $\Delta C_2 = \Ctwosq -\Ctwo^2$.  Now $\Ctwosq$ is the probability
of finding two sets of $2$-chains in $\Alex[p,q]$, with the possibility that some of the elements
can coincide. Let us call the points of these two chains $x, x'$ and $y,y'$, with $x\prec x'$ and $y
\prec y'$.  Thus, $\Ctwosq$ gets contributions from each type of coincidence. The first is simply
that there are no coincidences, i.e., that all 4 points $x,x',y,y'$ are distinct, which gives a
contribution $\Ctwo^2$.  Although this term $\sim \rho^4$, it cancels out in the expression for the
variance and therefore plays no role. The next type is the one-coincidence case. For this there are two
types: (i) $x=y'$ or $y=x'$, so that the two $2$-chains collapse to a single $3$-chain, and (ii)
$x=y$ or $x'=y'$, corresponding to the probability for a three element ``V'' or ``$\Lambda$'' shaped
causal set.  The contribution from (i) is clearly twice that of
$\Cthree$ whose $\rho$ dependence is $\sim \rho^3$, while the contribution from (ii) is
\begin{equation} 
2 \rho^3\negs \int\limits_{\Alex[p,q]} \negs dx \sqrt{-g(x)}\negs \int\limits_{J^+(x)\cap
  \Alex[p,q]}\negs \negs  \negs d x' \sqrt{-g(x')}  \negs \int\limits_{J^+(x)\cap \Alex[p,q]} \negs
\negs d y' \sqrt{-g(y')}
= 2 \rho   \int\limits_{\Alex[p,q]} \negs dx \langle C_1(x)
\rangle ^2 \sim \rho^3.   
\end{equation}   
Finally, there is a contribution from two-coincidences $x=y, x'=y'$, which is just $\Ctwo$ which
goes as $\sim \rho^2$. Thus, $\Delta C_2 \sim \rho^3$ or $\delta C_2 \sim \rho^{\frac{3}{2}}$, with the
dominant contribution coming from the one-coincidence case.  

In order to calculate the error for all our geometric parameters, we need to perform a similar
analysis for $k=3, 4$.  In each such case,  the dominant contribution to the error comes from the
one-coincidence case, since the no-coincidence contributions simply cancel out. 

\begin{figure}[ht] 
\centering \resizebox{5.5in}{!}{\includegraphics{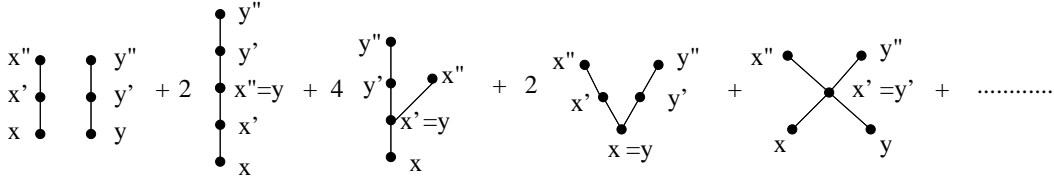}}
\vspace{0.5cm}
\caption{{\small The contributions to $\langle C_3^2 \rangle$ from no-coincidences and
    one-coincidence are shown.  }} \label{var.fig}
\end{figure}

For $k=3$, if $x\prec x' \prec x''$ and $y \prec y' \prec y''$, the no-coincidence case is again
simply $\langle C_3^2 \rangle$ and again cancels out.  The one-coincidence terms include:
(i) $x=y''$ or $y=x''$ which is  the $5$-chain, $\langle C_5
\rangle \sim \rho^5 $,  (ii) $x=y'$ or $y=x'$ or $x''=y'$ or $y''=x'$ which contribute 
\begin{equation} 
4 \rho^2  \int\limits_{\Alex[p,q]} \negs dy \sqrt{-g(y)} \negs \int\limits_{J^+(y) \cap \Alex[p,q]}
\negs dx \sqrt{-g(x)} \langle C_2(x) \rangle \langle C_1(x) \rangle \sim \rho^5,  
\end{equation}  
(iii) $x=y$ or  $x''=y''$ which contribute    
\begin{equation} 
2 \rho \int_{\Alex[p,q]} \negs dx \sqrt{-g(x)} \langle C_2(x) \rangle^2 \sim \rho^5,
\end{equation} 
and finally 
(iv) $x'=y'$ which contribute   
\begin{equation} 
\rho^3 \int_{\Alex[p,q]} \negs dx \sqrt{-g(x)}\int_{\Alex[p,q]} \negs dy \sqrt{-g(y)} \negs \negs \int\limits_{J^+(x)
  \cap J^+(y)\cap \Alex[p,q]} \negs \negs dx' \sqrt{-g(x')} \langle C_1(x)\rangle^2 \sim \rho^5. 
\end{equation} 
Figure \ref{var.fig} shows the contributions to $\langle C_3^2 \rangle$ from  the case of  no-coincidence and
the case of 1-coincidence. 
Since the no-coincidence term cancels out, $\delta C_3 \sim \rho^{\frac{5}{2}}$.  

A similar analysis shows that  for $\langle C_4^2 \rangle$ the no-coincidence term cancels out and the one-coincidence
cases  lead to a dependence  $\sim \rho^7$, so that    $\delta C_4 \sim \rho^{\frac{7}{2}}$ 
This generalises in a straightforward manner to 
\begin{equation} 
\delta C_k \sim \rho^{\frac{2k-1}{2}}. 
\end{equation} 

We are now in a position to calculate the dependence on $\rho$ of the errors in the $Q_k$ and
$S_k$: 
\begin{equation} 
\delta Q_k \sim  \frac{3}{k\rho^3} \Ck^{3/k-1} \delta C_k \sim \rho^{-1/2}.    
\end{equation} 
and 
\begin{equation}
 \delta S_k \sim  \frac{4}{k\rho^4} \Ck^{4/k-1} \delta C_k \sim \rho^{-1/2}.    
\end{equation} 
This immediately means that the errors $\delta T^{3n}$, $\delta R(0)$, $\delta R_{00}(0)$ go as
$\rho^{-1/2}$ and hence become smaller as $\rho$ increases.  The error in the dimension estimator 
is similarly given by 
\begin{eqnarray}
& &\delta n \biggl (  (4n+6)S_1 -3 (12n+10)S_2 + 3 (24n+14)S_3 -(40n+18)S_4 \biggr)     \nno \\ 
& = & - \biggl( (n+2)(2n+2)\delta S_1   - 3 (2n+2)(3n+2)\delta S_2 +  \nno \\ 
&& \qquad  3 (3n+2)(4n+2)\delta S_3  - (4n+2)(5n+2)\delta S_4 \biggr)   \nno \\ 
&& \Rightarrow \quad \delta n \sim  \rho^{-1/2}. 
\end{eqnarray}

\section{Conclusions and Remarks} 
\label{sseven} 

In this work we have found expressions for the proper time $T$ (Eqn (\ref{T})), the scalar
curvature $R(0)$ (Eqn (\ref{R})), the time-time component of the Ricci tensor $R_{00}(0)$ (Eqn
(\ref{Roo})), and a new dimension estimator (Eqn (\ref{dim})) from a causal set underlying a small causal diamond
$\Alex[p,q]$ in a generic spacetime in arbitrary dimensions. We find that the errors in these 
estimators goes as $\rho^{-1/2}$,  thus becoming  smaller as the sprinkling density is
increased, while keeping the volume of $\Alex[p,q]$ fixed.  Our results not only  verify the
deep relationship between order and Lorentzian geometry, but also provide new observables that can
be used to assess whether a  causal set is  manifold-like or not.   
  
Our calculation moreover brings to light an intriguing connection between two seemingly disparate
order theoretic structures in a causal set. While our expression for the scalar curvature is purely in
terms of the abundance of $k$-chains, the Benincasa-Dowker(BD)  scalar curvature $R_{BD}$ 
\cite{bd} is constructed from the abundance of $k$-inclusive intervals. In $4$-dimensions for example,
their expression for the scalar curvature is
\begin{eqnarray} 
\label{bd}
R_{BD}(0) = \frac{2}{6 \sqrt{\rho}} ( 1- (N_2(0)-9 N_3(0)+ 16N_4(0) -8 N_5(0)))
\end{eqnarray}  
where $N_k(0)$ are the number of $k$-inclusive intervals $I_k[x,0]$, where $x\ \prec \, \,
0$\footnote{Our notation differs from \cite{bd} where $k $ is replaced by $ k-1$,  in keeping with
our definition of a  $k$-chain.}.  As 
noted in Section \ref{stwo},  $k$-chains and $k$-inclusive intervals are very distinct order theoretic
structures. The continuum geometry however seems to link them via the scalar curvature. If such a
relationship exists, does it for example indicate manifold-likeness in a causal set? This and several related
questions remain to be investigated.  

Further comparisons between the two expressions for $R(0)$ are warranted.  Both contain alternating
sums, although the coefficients differ markedly. For one, the BD expression appears to have a strong
dimension dependence in the number of terms required -- for $n=2$ the sum is truncated at $k=4$,
while for $n=4$ it is truncated at $k=5$.  Our expression for $R(0)$ is in this sense independent of
$n$ -- it requires $\Cone,\Ctwo,\Cthree$ in {\it all} dimensions\footnote{As may be already evident
  to the astute reader, one could replace these three values $k=1,2,3$ with any other $k_1, k_2,
  k_3$, to get an expression for $R(0)$ in terms of $\langle C_{k_1} \rangle,\langle C_{k_2}
  \rangle,\langle C_{k_3} \rangle $. What is important is that this choice is not dimension
  dependent -- every choice works for every $n$.}. Moreover, the systematic determination of the coefficients for
$R_{BD}$ for arbitrary $n$ is fairly involved, whereas the coefficients in Eqn
(\ref{R}) and Eqn (\ref{U}) are simply the binomial coefficients $(-1)^k \binom{2}{k}$ and $(-1)^k
\binom{3}{k}$, respectively, in all dimensions. 

Apart from these, there are deeper differences in the two expressions for $R(0)$.  $R_{BD}(0)$ is
constructed from {\it all} inclusive intervals of the form $I_k[x,0]$ in the causal set for
$k=1,\ldots 5$. It is therefore an essentially non-local expression since it depends on the
structure of the causal set throughout a past (or future) neighbourhood of the element
$0$ in the causal set. The expression Eqn (\ref{R}) on the other hand is {\it only} valid in a small
causal diamond, and hence is strictly local and dependent on a proper choice of neighbourhood of the
element in the causal set. This means that the BD form for $R(0)$, being ``neighbourhood
independent'', can be readily used to obtain an action for the entire causal set, which has strong
implications for causal set theory. Unless the definition of a small neighbourhood can be made
entirely order theoretically in the causal set, our expression for $R(0)$ on the other hand cannot
be  used to obtain an action in a simple manner. Of course, in the {\it specific} case when the entire causal set is
approximated by a small causal diamond $\Alex[p,q]$ the action is simply
\begin{equation} 
S/\hbar =\sum_{s=1}^N R(e_s) = N R,  
\end{equation} 
where $e_s$ denotes an element in $\causet$,   since to this approximation $R(x)$ is the same
throughout $\Alex[p,q]$.   

Nevertheless, the geometric estimators we have found in this work take a big step towards the
order-Lorentzian geometry correspondence. We have a new dimension estimator which can determine the
manifold dimension for a curved spacetime with greater accuracy than the currently available flat
spacetime Myrheim-Meyer estimator. The local nature of the expression for the scalar curvature can
also be seen as an advantage since only a small neighbourhood of an element in the causal set is
required to determine the curvature, rather the than its entire past.  An obvious next step is to
follow up with a numerical analysis of causal sets which are approximated by different curved
spacetimes and to see how well our estimators work in these cases \cite{rsstwo}.

These observables can also be used as additional tests of manifold-likeness of a causal set. In
\cite{ss} the expectation values of a range of such observables, including the abundance $N_k$ of
$k$-element inclusive intervals was found from Monte Carlo simulations of 2-d causal set quantum
gravity. Comparing with the flat spacetime distribution of the $N_k$, these observables were used to
demonstrate that the dominant contribution to the causal set path integral in 2d comes from flat
spacetime suggesting  that manifold-like behaviour is emergent. The estimators we have obtained in this
current work could well be employed to calculate more accurately how close to flatness one is -- is
the 2-d universe just a little positively curved, for example?

The extended hope of the current analysis is also that as more of the order theoretic basis of
geometry is uncovered it may be possible to find a meaningful order theoretic definition of locality
which translates to our commonly held (albeit Riemannian geometry based) notions of locality in the
continuum.

\section{Appendix}  
\label{One.ap} 

To evaluate the $\Ck$ we find it  convenient to transform from the Cartesian coordinates $x^i$ to
spherical polar coordinates $x^i=r f^i(\Omega)$, where
\[  
f^i(\theta_1, \theta_2 \ldots \theta_{n-2}) =
\left\{
              \begin{array}{ll}
                   \prod\limits_{k=1}^{n-i-1}\sin\theta_k \cos\theta_{n-i}  & (i>1)\\ \\ 
                   \prod\limits_{k=1}^{n-2}\sin\theta_k & (i=1)
              \end{array}
\right.
\]
It is also useful to express the radial coordinate for any $x_1 \in \Alex[p,q]$  in light-cone
coordinates: 
\begin{eqnarray} 
 r_1 & = & \frac{v_1-u_1}{2} = \frac{1}{2} \biggl( \biggl(\frac{T}{2}-u_1\biggr) - \biggl(
\frac{T}{2} -v_1\biggr)\biggr) \nonumber \\ 
\Rightarrow \quad r_1^l &= &\frac{1}{2^l}\sum\limits_{k=0}(-1)^k \binom{l}{k}
\biggl(\frac{T}{2}-v_1\biggr)^k\biggl(\frac{T}{2}-u_1\biggr)^{l-k}  \nonumber 
\end{eqnarray} 
Since the metric $\etalc$ in $\Alexflat$ in light cone coordinates is  
\begin{equation}
ds^2 = -du dv + \biggl( \frac{v-u}{2}\biggr)^2 d \Omega^2,  
\end{equation} 
the associated measure of any integral over $\Alexflat$ is 
\begin{eqnarray} 
\sqrt{-\etalc} \negs &=& \negs \frac{1}{2}\biggl(\frac{v-u}{2}\biggr)^{n-2}  \biggl(\prod_{i=1}^{n-3}
\sin^{(n-i-2)}\theta_i\biggr) \nonumber \\ 
&=&\negs \frac{1}{2^{n-1}}\biggl(\prod_{i=1}^{n-3}
\sin^{(n-i-2)}\theta_i\biggr)
\sum\limits_{k=0}^{n-2} (-1)^k \binom{n-2}{k}\biggl(\frac{T}{2}-v\biggr)^k
\biggl(\frac{T}{2}-u\biggr)^{n-2-k}.  
\end{eqnarray} 
Finally,  we will find it useful to define   the following pair of integrals: 
\begin{eqnarray} 
\intL(\al,v,T)  &= & \int\limits_{-\frac{T}{2}}^{v} du\biggl(\frac{T}{2}-u\biggr)^\al = \frac{1}{\al+1}
\biggl( T^{\al+1} - \biggl(\frac{T}{2}-v\biggr)^{\al+1}\biggr) \nonumber \\ 
\intL(\al,\be,T) &=&   \int\limits_{-\frac{T}{2}}^{\frac{T}{2}} dv \biggl(\frac{T}{2}-v\biggr)^\be
\intL(\al,v,T) = \frac{T^{\al+\be+2}}{(\be+1)(\al+\be+2)}. 
\end{eqnarray}

The above identities will now be used to evaluate the set of integrals
$\intI_{1,2,3,4}(m)$ required for  the calculation of $\Ck$ (Eqns (\ref{ctwo}), (\ref{cthree})
(\ref{ckplusone})). 

\begin{enumerate}  
\item  \begin{eqnarray}
\label{intone} 
\intI_1(m)&=&\int dx_1 T_1^m \nonumber \\ 
&=&  \int\limits_{-\frac{T}{2}}^{\frac{T}{2}} dv_1 \int\limits_{-\frac{T}{2}}^{v_1} du_1 \int
d\Omega \sqrt{-
\eta(u_1,v_1)} \, \, T_1^m \nonumber  \\ 
&=&  \frac{A_{n-2}}{2^{n-1}} \sum\limits_{k=0}^{n-2} \binom{n-2}{k}
\intL(\frac{m}{2}+n-k-2,\frac{m}{2}+k,T)  \nonumber \\ 
&=&  \frac{A_{n-2}T^{n+m}}{2^{n-1}}\frac{(n-2)!(\frac{m}{2})!}{(n+m)(n+\frac{m}{2}-1)!}.
\end{eqnarray} 

\item For $m+2$ the above equation gives us the relation 
\begin{equation} 
\label{ione} 
\intI_1(m+2) = \intI_1(m) T^2 \frac{(n+m)(\frac{m}{2}+1)}{(n+m+2)(n+\frac{m}{2})}= \intI_1(m)T^2 g_1(m).
\end{equation}

\item 
As noted in Section \ref{sthree} any integral of the form 
\begin{equation} 
I_2(m) = \int\limits_{\Alexflat}  dx _1 (x_1^i)^2 T_1^m 
\end{equation} 
is independent of the spatial direction $i$ because of the spatial symmetry of $\Alexflat$. Thus,  we can  choose 
$x^{n-1}=r \cos \theta_1$ to simplify our calculation so that  
\begin{eqnarray} 
\label{itwo} 
\intI_2(m) &=& \int dx_1T_1^m r_1^2 \cos^2\theta_1 \nonumber \\ 
&=& \int\limits_{-\frac{T}{2}}^{\frac{T}{2}} dv_1 \int\limits_{-\frac{T}{2}}^{v_1} du_1 \int\ d\Omega
  \frac{\cos^2 \theta_1}{2^{n+1}} 
  \biggl( \frac{T}{2}-u_1 \biggr)^{\frac{m}{2}}    \biggl( \frac{T}{2}-v_1 \biggr)^{\frac{m}{2}}
  \biggl(v_1-u_1 \biggr)^n \nonumber \\ 
&=& \frac{A_{n-2}}{2^{n+1}}\frac{T^{n+m+2}}{(n-1)(n+m+2)}\frac{n!
  (\frac{m}{2})!}{(n+\frac{m}{2}+1)!}\nonumber \\ 
&=&\intI_1(m)T^2 \frac{n(n+m)}{4(n+m+2)(n+\frac{m}{2}+1)(n+\frac{m}{2})} \nonumber \\
&=& \intI_1(m)T^2 f_2(m).
\end{eqnarray} 
The next integral can be split into three parts
\begin{equation} 
\label{ithree} 
\intI_3(m)= \int dx_1 t_1^2 T_1^m =\intI_3^a + \intI_3^b+\intI_3^c 
\end{equation} 
where, using   $\al_k=n-2-k+\frac{m}{2}, \be_k=k+\frac{m}{2}$ 
\begin{eqnarray} 
\intI_3^a(m)&=&=\frac{1}{4} \int dx_1 u_1^2T_1^m \nno \\ 
&=& \frac{A_{n-2}}{2^{n+1}}\sum\limits_{k=0}^{n-2}(-1)^k \binom{n-2}{k}\biggl( \frac{T^2}{4}
  \intL(\al_k,\be_k,T)-T\intL(\al_k+1,\be_k,T)+\intL(\al_k+2,\be,T)\biggr),  \nno \\ 
\end{eqnarray} 
\begin{eqnarray}
\intI_3^b(m) &=&\frac{1}{4}  \int dx_1 v_1^2T_1^m \nno \\
&=& \frac{A_{n-2}}{2^{n+1}}\sum\limits_{k=0}^{n-2}(-1)^k \binom{n-2}{k}\biggl( \frac{T^2}{4}
  \intL(\al_k,\be_k,T)-T\intL(\al_k,\be_k+1,T)+\intL(\al_k,\be+2,T)\biggr),  \nno \\ 
\end{eqnarray} 
and 
\begin{eqnarray} 
\intI_3^c(m) &=&\frac{2}{4}  \int dx_1 u_1 v_1 T_1^m \nno \\
&=& \frac{A_{n-2}}{2^{n}}\sum\limits_{k=0}^{n-2}(-1)^k \binom{n-2}{k}\biggl( \frac{T^2}{4}
  \intL(\al_k,\be_k,T) -\frac{T}{2}\intL(\al_k+1,\be_k,T) \nno \\ 
&&  -\frac{T}{2}\intL(\al_k,\be_k+1,T)  +\intL(\al_k+1,\be+1,T)\biggr).  
\end{eqnarray} 
After some algebra, we find that 
\begin{equation}
 \intI_3(m)=\intI_1(m)T^2\frac{8n+m(m+2)(m+n+2)}{4(2+m+n)(2n+m)(2n+m+2)}=\intI_1(m)T^2 f_3(m).  
\end{equation}

\item 

Finally, 
\begin{eqnarray}
\intI_4(m)&=&\int dx_1 \biggl(\frac{T}{2}-t_1 \biggr)^2 T_1^m   \nno \\ 
&=& \frac{T^2}{4} \intI_1(m)+\intI_3(m) - T \tilde \intI_4(m). 
\end{eqnarray} 
Evaluating 
\begin{eqnarray} 
\tilde\intI_4(m) &=& \int dx_1 t_1 T^m \nno \\ 
&=& \intI_1(m) 
\frac{T}{2}\biggl(1-\frac{m+n}{m+n+1}-\frac{(\frac{m}{2}+1)(m+n)}{(m+n+1)(\frac{m}{2}+n)} \biggr)
\nno \\ 
&=&  \intI_1(m) T \tilde f_4(m).   
\end{eqnarray} 
We can then express 
\begin{equation}
\label{ifour} 
\intI_4(m) = \intI_1(m) T^2  f_4(m)
\end{equation} 
where $f_4(m)=\frac{1}{4} + \tilde f_4(m)+f_3(m)$.  
\end{enumerate} 

Gathering these expressions 
\begin{eqnarray}
\label{gfs} 
g_1(m)= \frac{(n+m)(\frac{m}{2}+1)}{(n+m+2)(n+\frac{m}{2})} & \quad & f_2(m) =
\frac{n(n+m)}{4(n+m+2)(n+\frac{m}{2}+1)(n+\frac{m}{2})} \nno \\   
f_3(m) =\frac{8n+m(m+2)(m+n+2)}{4(2+m+n)(2n+m)(2n+m+2)} &\quad &  f_4(m) =  \frac{1}{4} - \tilde f_4(m)+f_3(m). 
\end{eqnarray}

\end{document}